\documentclass[
 reprint, 
 amsmath, amssymb, 
 aps, 
]{revtex4-2}

\usepackage{graphicx}  
\usepackage{amssymb}   
\usepackage{dcolumn}
\usepackage{amsmath}   
\usepackage{bm}		   
\usepackage[a]{esvect} 

\newcommand{\vect}[1]{\bm{#1}} 
\newcommand{\pp}[1]{\partial #1} 
\newcommand{\priv}[2]{\frac{\pp{#1}}{\pp{#2}}} 

\newcommand{\rbrac}[1]{\left( #1 \right)}
\newcommand{\sbrac}[1]{\left[ #1 \right]}

\newcommand{\ie}{\textit{i.e.},\,}
\newcommand{\eg}{\textit{e.g.},\,}

\newcommand{\viz}{\textit{viz.}\,}
\newcommand{\ts}{\mathrm{s}}

\newcommand{\tkin}{\mathrm{kin}}
\newcommand{\tmag}{\mathrm{mag}}
\newcommand{\texp}{\mathrm{exp}}
\newcommand{\tsat}{\mathrm{sat}}
\newcommand{\tnl}{\mathrm{nl}}

\newcommand{\tsc}{\mathrm{sc}}
\newcommand{\tcrit}{\mathrm{crit}}
\newcommand{\tend}{\mathrm{end}}
\newcommand{\tres}{\mathrm{res}}
\newcommand{\tmax}{\mathrm{max}}
\newcommand{\Reh}{\mbox{Re}}
\newcommand{\Rem}{\mbox{Rm}}

\newcommand{\Prm}{\mbox{Pm}}
\newcommand{\Mach}{\mathcal{M}}

\newcommand{\nquad}[1][1]{\hspace*{#1em}\ignorespaces}

\newcommand{\aref}[1]{\hyperref[#1]{Appendix~\ref{#1}}}

\begin{document}

\preprint{APS/123-QED}

\title{The universal growth of magnetic energy during the nonlinear \\ phase of subsonic and supersonic small-scale dynamos}

\author{Neco Kriel}
\email{neco.kriel@anu.edu.au}
\affiliation{Research School of Astronomy and Astrophysics, Australian National University, 233 Mount Stromlo Road, Stromlo ACT 2612, Australia}

\author{James R. Beattie}
\email{james.beattie@princeton.edu}
\affiliation{
Canadian Institute for Theoretical Astrophysics, University of Toronto, Toronto, M5S3H8, ON, Canada \\
Department of Astrophysical Sciences, Princeton University, Princeton, 08540, NJ, USA 
}

\author{Mark R. Krumholz}
\affiliation{Research School of Astronomy and Astrophysics, Australian National University, 233 Mount Stromlo Road, Stromlo ACT 2612, Australia}

\author{Jennifer Schober}
\affiliation{Argelander-Institut f\"ur Astronomie, Universit\"at Bonn, Auf dem H\"ugel 71, 53121 Bonn, Germany}

\author{Patrick J. Armstrong}
\affiliation{Research School of Astronomy and Astrophysics, Australian National University, 233 Mount Stromlo Road, Stromlo ACT 2612, Australia}

\begin{abstract}
    Small-scale dynamos (SSDs) amplify magnetic fields in turbulent plasmas. Theory predicts nonlinear magnetic energy growth $E_\mathrm{mag} \propto t^{p_\mathrm{nl}}$, but this scaling has not been tested across flow regimes. Using a large ensemble of SSD simulations spanning subsonic to supersonic turbulence, we measure linear growth ($p_\mathrm{nl} = 1$) in subsonic flows and quadratic growth ($p_\mathrm{nl} = 2$) in supersonic flows. In all cases, the nonlinear dynamo converts a nearly constant fraction $\sim 1/100$ of the turbulent kinetic energy flux into magnetic energy, and the nonlinear phase has a characteristic duration $\Delta t \approx 20\,t_0$, where $t_0$ is the outer-scale turnover time. By isolating the onset of magnetic backreaction in SSDs, our statistical ensemble approach identifies a robust efficiency and duration for the nonlinear SSD that can be used to interpret more complex astrophysical and laboratory plasmas.
\end{abstract}

\maketitle

\section{Introduction}

    Small-scale dynamo (SSD) action describes the process by which motions in a plasma with dynamically weak magnetic fields amplify and then maintain the magnetic energy density, $E_\tmag$, to levels comparable to the turbulent kinetic energy density, $E_\tkin$ \citep{brandenburg2012current, rincon2019dynamo, rempel2023small}. This process is ubiquitous across astrophysical, geophysical, and laboratory environments, both those where the plasma supports turbulent subsonic motions (turbulent sonic Mach number $\Mach = u_0/c_s < 1$, where $u_0$ is the rest-frame root-mean-squared velocity and $c_s$ the sound speed) and where they are supersonic ($\Mach > 1$). The fields generated in this process magnetize the plasma between galaxies \citep{Roh2019_galaxy_clusters, tevlin2025magnetic}, provide pressure support against collapse in galaxy mergers \citep{whittingham2021impact}, enable rapid spin-down of stellar merger remnants \citep{palenzuela2022turbulent, ryu2025magnetic}, and modify cosmic ray propagation via curvature acceleration in the interstellar medium (ISM) \citep{kempski2023cosmic, Lemoine2023_curvature_accleration}. Direct evidence for SSD action comes both from \textit{in situ} observations in the Earth's subsonic magnetosheath \citep{voros2025turbulent} and from supersonic laboratory laser experiments \citep{Meinecke2014_SSD_lab_Nature, tzeferacos2018laboratory, Bott2021_supersonic_dynamo_in_lab, Bott2021_time_resolved_lab_dynamo}.

    SSDs evolve through a number of distinct phases \citep{moffatt1970dynamo, vainshtein1972origin, kulsrud1992spectrum, schekochihin2002model, schekochihin2004simulations, maron2004nonlinear, rincon2019dynamo}. At first the magnetic field is too weak to exert significant forces on the plasma and thus the velocity field is kinematic, which leads to exponential amplification of the magnetic energy density, $E_\tmag \propto \exp(\gamma_\texp t)$. Here $\gamma_\texp$ is the kinematic growth rate (with units of inverse time), and the dimensionless product $\gamma_\texp t_0$ measures how efficiently kinetic energy is converted into magnetic energy over one outer-scale eddy turnover time $t_0 = \ell_0/u_0$, where $u_0$ is the flow velocity on the outer scale $\ell_0$ \citep{moffatt1970dynamo, kulsrud1992spectrum}. For magnetic Prandtl numbers $\Prm \gtrsim 1$, the dimensionless growth rate is predicted to scale as $\gamma_\texp t_0 \propto \Reh^{1/2}$ for $\Mach < 1$ and $\gamma_\texp t_0 \propto \Reh^{1/3}$ for $\Mach > 1$, where $\Reh \sim u_0 \ell_0 / \nu$ is the hydrodynamic Reynolds number \citep{schober2012magnetic} and $\nu$ is the kinematic viscosity. Field amplification is driven by chaotic stretching \citep{rincon2019dynamo}, which occurs on an eddy turnover timescale $t_\ell \sim \ell/u_\ell$, where $u_\ell$ is the flow speed on size scale $\ell$, and thus is fastest on the viscous scale $\ell_\nu$, which, for Kolmogorov turbulence, is the scale on which eddies evolve on the shortest timescales \citep{bian2019decoupled, brandenburg2019reversed, grete2021matter, kriel2022fundamental, kriel2025fundamental, beattie2025taking}. However, once the kinematic dynamo amplifies the field such that $E_\tmag \approx E_\tkin$ on scale $\ell_\nu$, the magnetic field begins to back react on the velocity, marking the end of the kinematic phase and the start of the nonlinear phase. During this phase the dominant stretching scale $\ell_s$ shifts to larger $\ell$ with longer $t_\ell$, driving a slower, secular dynamo. As the field grows, it suppresses stretching on ever-larger scales, until $\ell_s$ reaches a maximum scale. At this point the dynamo reaches the saturated phase, and becomes statistically steady, driven MHD turbulence \citep{beattie2025spectrum}.

    Both numerical simulations \citep{schekochihin2004simulations, kriel2022fundamental, kriel2025fundamental}, and laboratory experiments \citep{tzeferacos2018laboratory, Bott2021_supersonic_dynamo_in_lab, Bott2021_time_resolved_lab_dynamo} confirm that the evolution of $E_\tmag$ during the kinematic phase agrees well with both the Kazanstev, Anderson \& Kulsrud SSD theory \citep{Kazantsev1968, kulsrud1992spectrum} as well as with numerical works in the $\Mach > 1$ regime \citep{Federrath2011_supersonic_dynamo, Federrath2014_supersonic_dynamo, kriel2025fundamental}. However, this well-understood phase likely ends almost immediately in real astrophysical plasmas. This is because the growth rate is $\gamma_\texp t_0 \propto \Reh^{1/2}$, and many warm or cold astrophysical plasmas, at least in our Galaxy and its surrounding medium, often have $\Reh \sim 10^3-10^{10}$, thus making the kinematic phase persist for only a small fraction of the outer dynamical timescale. Hence, almost any observable astrophysical system where the field is growing will instead be in the nonlinear phase, which is a regime that has not yet been explored with the same level of numerical detail as the kinematic dynamo.
    
    While there is indication that the nonlinear phase yields secular growth for the magnetic energy \citep{beresnyak2012universal, cho2009growth, seta2020seed, galishnikova2022tearing, kriel2025fundamental}, $E_\tmag = \alpha_\tnl t^{p_\tnl}$, there are no simulations to date that have measured the efficiency $\alpha_\tnl$ and growth order $p_\tnl$ across a wide range of plasma Reynolds number $\Reh$ and Mach number $\Mach$. For incompressible Kolmogorov-like turbulence, where $E_\tkin(\ell) \propto \ell^{5/3}$, the prevailing expectation and hint from numerical simulations is that $p_\tnl = 1$ \citep{kulsrud1992spectrum, schekochihin2002model, cho2009growth, beresnyak2012universal, xu2016turbulent}, and $\alpha_\tnl \propto \varepsilon$, where $\varepsilon \sim \rho_\ell u_\ell^3/\ell$ is the turbulent energy flux through scale $\ell$, with $\rho_\ell$ the density associated with motions on that scale (which is constant in the incompressible limit). In contrast, for highly-compressible, shock-dominated turbulence, where $E_\tkin(\ell) \propto \ell^2$ (as in Burgers-like turbulence \citep{Federrath2013_universality, Federrath2021_the_sonic_scale, kriel2025fundamental, beattie2025taking}), the outcome is debated: some models \citep{schleicher2013small, schober2015saturation} predict quadratic growth ($p_\tnl = 2$), while others suggest universal linear growth ($p_\tnl = 1$) \citep{beresnyak2012universal}.

    This lack of consensus about the nonlinear growth order $p_\tnl$ and efficiency $\alpha_\tnl$, particularly in highly-compressible turbulence, is largely due to a dearth of numerical and experimental guidance, which is missing in large part because it is extremely challenging to accurately measure the nonlinear phase. At the modest values of $\Reh$ accessible to numerical simulations and laboratory experiments, the nonlinear phase is hard to identify, due in part to the strong fluctuations of integral quantities that can easily mask underlying trends, especially in supersonic flows. Our goal in this study, is to take on these challenges by focusing on plasmas with $\Prm = 1$, for which the nonlinear phase we resolve, can only be the backreaction stage that begins once $E_\tmag \approx E_\tkin$ on the viscous scale. This choice also maximizes the computational dynamic range over which the nonlinear backreaction can operate, because the viscous and resistive dissipation scales for any choice of $\Reh$ or $\Rem$ coincide with one-another ($\ell_\nu \approx \ell_\eta$). Within this setting we directly measure $\alpha_\tnl$ and $p_\tnl$ across a broad range of plasma parameters (in $\Reh$ and $\Mach$), allowing us, for the first time, to confront theoretical predictions for the backreaction during the nonlinear phase with highly-detailed measurements from simulations. At the same time, our results provide a well-controlled reference point for future studies of more realistic hot astrophysical plasmas with $\Prm \gg 1$, where this backreaction stage may be preceded by an additional nonlinear, secular growth phase in the sub-viscous range (see the discussion in the Conclusion).

\section{Numerical Methods}

\subsection{Numerical Simulations} \label{sec:methods:sims}
    
    \begin{figure}
        \centering
        \includegraphics[width=\linewidth]{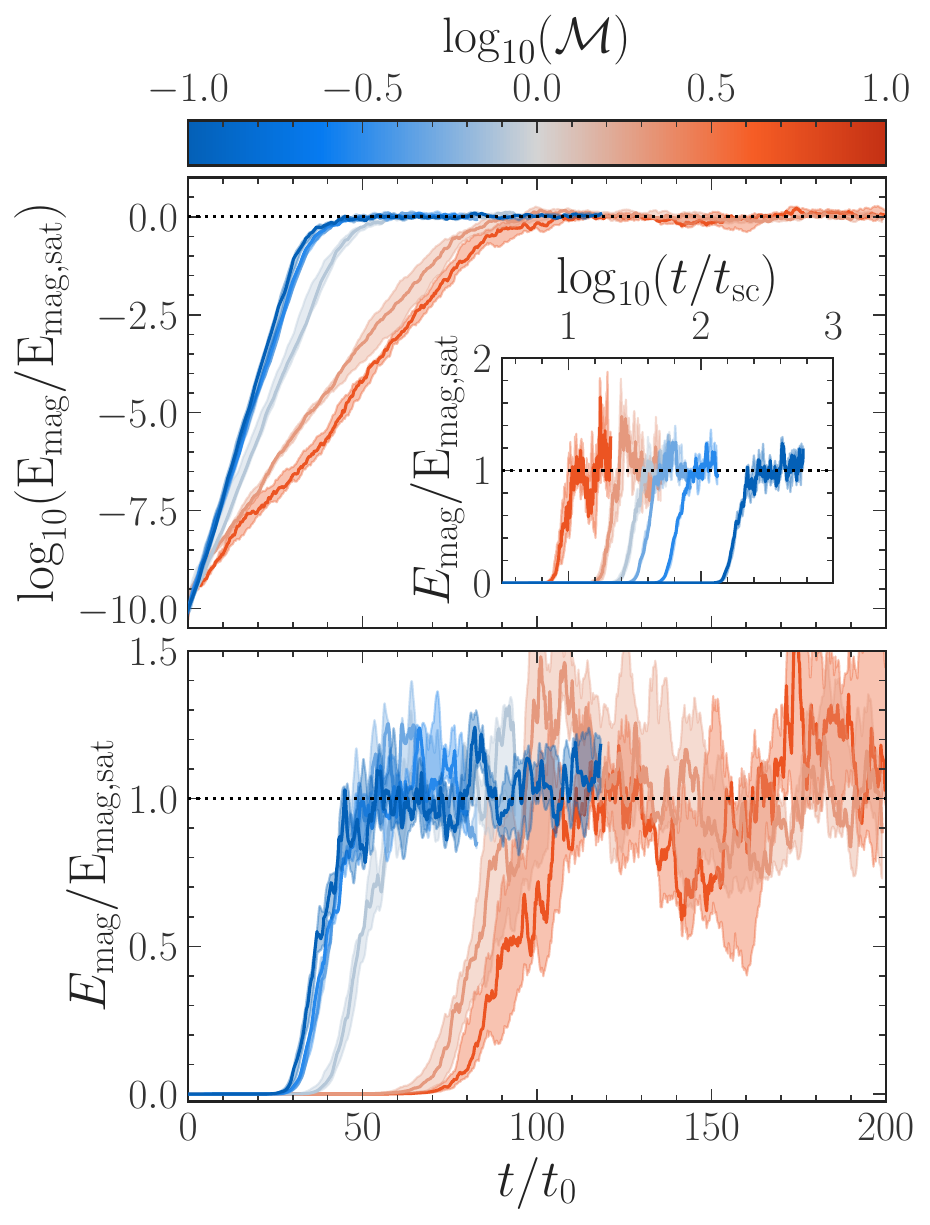}
        \caption{Time evolution of the volume-averaged magnetic energy $E_\tmag$, normalized by the final saturated value $E_{\tmag,\tsat}$, for a subset of SSD simulations with $\Reh = 1500$. The top panel shows the energy in logarithmic scale, and the bottom panel (and inset) shows it in linear scale. Each simulation configuration is repeated five times with independent random seeds for the turbulent forcing; we plot the median trend across realizations as a solid line, with the 16th to 84th percentile range indicated by a shaded band. Colors indicate the average sonic Mach number $\Mach$ of each configuration. In the main axes, time is shown in units of the turbulent turnover time on the outer scale, $t_0$, and rescaled to the sound-crossing time, $t_\tsc$, in the inset.}
        \label{fig:time_series}
    \end{figure}

    We used a modified version of the \textsc{flash} code \citep{Fryxell2000, Dubey2008, Waagan2011, Federrath2021_the_sonic_scale} to run a series of direct numerical simulations that solve the compressible, non-ideal MHD equations for an isothermal plasma in a three-dimensional periodic box \citep{kriel2025fundamental}; the full set of equations we solve is provided in End Matter. We work in dimensionless units where the box size $L$, mean density $\rho_0$, $c_s$, and mean thermal energy $\rho_0 c_s^2$ are unity. The natural unit of time for this system is the sound-crossing time, $t_\tsc = L / c_\ts = 1$, and the turbulent velocity $u_0$ directly sets the sonic Mach number, $\Mach = u_0/c_s$. By varying $u_0$, we control the speed of the flow relative to $t_\tsc$, and in turn can explore subsonic ($t_\tsc < t_0$) and supersonic ($t_0 < t_\tsc$) regimes. We use the \textsc{turbgen} \citep{Federrath2022_TurbGen} forcing-module to drive purely-solenoidal ($\vect{\nabla}\cdot\vect{f}=0$) turbulence with momentum source term $\vect{f}$, following the standard protocol outlined in \citet{kriel2022fundamental}, allowing us to set $\Mach \in [5\times 10^{-2}, 5]$ (see End Matter for more details).
        
    By varying the (constant in space and time) kinematic viscosity coefficient, $\nu$, across different simulation configurations, we explore a range of hydrodynamic Reynolds numbers, $\Reh \in [10^3, 5\times 10^3]$, where we set the magnetic Prandtl number $\Prm \equiv \nu/\eta = \Rem/\Reh = 1$; $\eta$ is the magnetic resistivity coefficient and $\Rem$ is the magnetic Reynolds number. We quote $\Reh$ and $\Rem$ using $u_0$, which is the rest-frame root-mean-squared velocity measured in the statistically steady kinematic phase. For convenience, we use these kinematic values to label simulations and compare across the parameter space. During the nonlinear and saturated phases, magnetic backreaction slightly suppresses the velocity amplitude, so the instantaneous effective values of $\Reh$ and $\Rem$ may be smaller than their nominal kinematic values. This effect, however, is minor for $\Prm = 1$, becoming more prominent for $\Prm \gg 1$ \citep{seta2021saturation, shivakumar2025numerical}. Because $\Reh = \Rem \gtrsim \Rem_\tcrit \approx 100$, all simulations are above the critical $\Rem$ required to undergo all phases of the SSD \citep{schekochihin2007fluctuation, schober2015saturation, Seta2020_critical_Rm}. In total we consider $12$ distinct (but over many different statistical realizations of $\vect{f}$, detailed in the next paragraph) combinations of $\Mach$ and $\Reh$ -- see End Matter for a complete list of simulations. We also ran pilot simulations at lower $\Reh$, but found that the volume-integrated kinetic and magnetic energies had relative fluctuations that were too large for our analysis. When the viscous and resistive dissipation scales $\ell_\nu$ and $\ell_\eta$ lie close to the forcing scale $\ell_0$, the cascade is truncated and fluctuations remain confined to large scales. The effective number of independent eddies contributing to volume averages, $N_{\rm eddies} \sim (L/\ell_\nu)^3$, is then small, so the fractional fluctuations $\propto N_{\rm eddies}^{-1/2}$ remain large in the nonlinear phase. For this reason we adopt moderately turbulent flows with $\Reh \gtrsim 10^3$ as our practical lower bound. Finally, we caution readers that there are two different conventions in use in the literature to define the Reynolds number; our definition, $\Reh \sim u_0 \ell_0 / \nu$, is a factor of $2\pi$ larger than that used by some authors \citep[\eg][]{brandenburg2023dissipative, galishnikova2022tearing}, so in the convention used by these authors, our simulations span $\Reh \in [1.5\times 10^2, 8\times 10^2]$.
    
    In all our simulations we initialize a weak seed magnetic field $E_{\tmag, 0} \equiv E_\tmag(t=0) = 10^{-10} E_\tkin$, where $E_\tkin$ is the kinetic energy once turbulence is fully established in the kinematic phase. We evolve each simulation instance long enough to unambiguously identify the onset and transition out of the nonlinear phase (see next section). To reduce the impact of fluctuations during the nonlinear phase on our fit parameters, we repeat each configuration at least five times with different random seeds for $\vect{f}$. We perform runs at three resolutions, $N_\tres \in \{288^3, 576^3, 1152^3\}$, and run a total of $89$ independent simulation instances. We plot the time evolution for the integral magnetic field energy of a representative subset of our simulations ($\Mach$-varied with $\Reh = 1500$) in Fig.~\ref{fig:time_series}, where the classical SSD phases are clear to see. The main panels use $t_0$ to emphasize the evolution in outer-scale turnover times, while the inset re-normalizes the same curves in units of the sound-crossing time $t_\tsc$, highlighting how the ordering of the nonlinear phase onset changes when measured against $t_\tsc$ rather than $t_0$.

\subsection{Fitting Procedure} \label{sec:methods:fitting}

    To constrain growth timescales during the nonlinear phase, we begin by binning the raw time series of volume-averaged magnetic energy density $E_\tmag$, for each of our simulations, into intervals of $t_0$. Within each time bin $t_i$, we compute the mean $\mu_i$ and standard deviation $\sigma_i$ of both $E_\tmag$ and $\ln(E_\tmag)$ in bin $i$; $\mu_i^{(\ln)}$ and $\sigma_i^{(\ln)}$ indicate the mean and standard deviation of $\ln(E_\tmag)$ in bin $i$. This yields, for each simulation $j$, a dataset $\vect{\mathcal{D}}_j \equiv (t_i, \mu_i, \mu_i^{(\ln)}, \sigma_{i}, \sigma_i^{(\ln)})$, where data points are (approximately) uncorrelated, given that the turbulence correlation time is $t_0$.
    
    Using a hierarchical two-stage strategy, we then employ a Bayesian approach to fit each $\vect{\mathcal{D}}_j$ to models for the time evolution of magnetic energy. In the first stage we only constrain the exponential and saturated behavior, and in the second stage we fit the full three-phase model with priors informed by the first. This ensures that the more complex nonlinear-phase dynamics remain stable and well-constrained, especially in our supersonic simulations, where large fluctuations can make the nonlinear phase difficult to identify. Anchoring in this way also improves the sampling efficiency.
    
    Our first stage models only the kinematic and saturated phases, approximating the transition between these phases as instantaneous,
    \begin{align}
        f_1(t \mid \vect{\theta}_1) =
            \begin{cases}
                E_0 \exp(\gamma_\texp t) , & t < t_\tsat \\
                E_0 \exp(\gamma_\texp t_\tsat) , & t \geq t_\tsat
            \end{cases}
    \end{align}
    with $\vect{\theta}_1 = (E_0, \gamma_\texp, t_\tsat)$, and the second stage models all dynamo phases self-consistently as 
    \begin{align}
        f_2(t \mid \vect{\theta}_2) =
            \begin{cases}
                E_0 \exp(\gamma_\texp t) , & t < t_\tnl \\
                E_0 \exp(\gamma_\texp t_\tnl) \\
                \nquad[2] + \alpha_\tnl (t - t_\tnl)^{p_\tnl} , & t_\tnl \leq t < t_\tsat \\
                E_\tsat , & t \geq t_\tsat
            \end{cases} \label{eqn:integral_energy_model}
    \end{align}
    where $\vect{\theta}_2 = (E_0, E_\tsat, \gamma_\texp, t_\tnl, t_\tsat, p_\tnl)$. Here, $\alpha_\tnl$ is not a fit parameter in the second stage, because it is implicitly set by the requirement that $f_2$ must remain continuous at $t = t_\tsat$.
    
    For our fits in both stages, we assume that each data point $d_i$ in the time series is an independent sample from a Gaussian distribution around a mean trend, so for any given time series of simulation data $\vect{\mathcal{D}}_j$, the log likelihood function is
    \begin{align}
        \ln \mathcal{L}(\vect{\theta}_s \mid \vect{\mathcal{D}}_j)
            = -\frac{1}{2} \sum_{i} \sbrac{
                \frac{ \left( d_i - f_s(t_i \mid \vect{\theta}_s)\right)^2 }{ e_i^2 }
            } , 
    \end{align}
    where we fit $(d_i, e_i) = (\mu_i^{(\ln)}, \sigma_i^{(\ln)})$ in stage one ($s = 1$), and we fit $(d_i, e_i) = (\mu_i, \sigma_i)$ in stage two ($s = 2$). This is because, as is apparent from Fig.~\ref{fig:time_series}, it is much easier to identify the kinematic (exponential growing) phase in logarithmic space, and the nonlinear (backreaction) phase in linear space.
    
    We fit using a Markov Chain Monte Carlo method \cite{Foreman-Mackey13a}; see End Matter for details on our sampling parameters. In the first stage we adopt uniform (unbiased) priors on $\gamma_\texp \in (0, 10]$ and $t_\tsat \in (0, t_\tend)$, where $t_\tend$ is the final time bin, and a log-uniform prior on $E_0 \in [10^{-30}, 10^{-5}]$ (due to its wide dynamic range). In the second stage, the \textit{priors} for $E_0$, $\gamma_\texp$ and $E_\tsat$ are taken to be the \textit{posteriors} from stage one. We assign a uniform prior $p_\tnl \in [1,2]$, consistent with existing theories, while, for $t_\tnl$ and $t_\tsat$ we use priors that are uniform on the intervals $[0,t_{\tnl, \tmax}]$ and $[t_\tnl,t_{\tsat,\tmax}]$, where $t_{\tnl, \tmax}$ is the time bin where $dE_\tmag/dt$ is its maximum, and $t_{\tsat,\tmax}$ is the earliest time $t_i$ for which $dE_\tmag/dt < 0$.

    After fitting each individual simulation, we aggregate posteriors across runs with the same resolution and plasma parameters. This is the key step in our procedure, where we average out the inevitably large fluctuations in any single realization. For a given configuration, all realizations are statistically equivalent and differ only in their random driving, so we combine their posterior samples with equal weighting. Fitting each simulation separately preserves the phase structure of each realization, while still averaging over ensemble variability. This is crucial, because identical realizations may enter the nonlinear and saturated phases at slightly different times relative to one another. Accurately constraining these transition times is essential for robustly measuring the nonlinear growth dynamics, which would be averaged out if all instances of a configuration were fit simultaneously. All the results we present in the remainder of this study are derived from these aggregated samples; percentiles of these derived quantities are reported in Table~\ref{table:summary}, and the raw data, along with our implementation of the routines discussed here, are publicly available online \citep{mydata}.

\section{Results}

\subsection{Kinematic dynamo growth}

    \begin{figure}
        \centering
        \includegraphics[width=\linewidth]{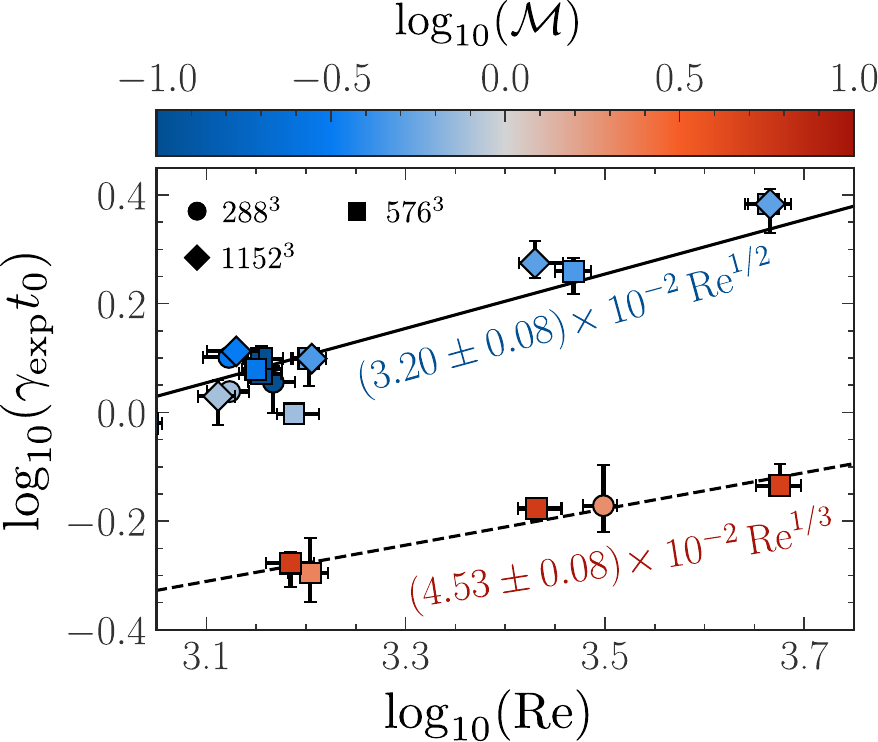}
        \caption{Kinematic growth rates, $\gamma_\texp$, as a function of plasma Reynolds number, $\Reh$, with markers indicating different $N_\tres$, and colors showing $\Mach$. Each data point shows the median marginal posterior probability for $\gamma_\texp$ derived by fitting Eqn.~\ref{eqn:integral_energy_model} to each plasma combination, $(N_\tres,\Mach,\Reh)$. The vertical error bars show the 16th to 84th percentile range from the fits and the horizontal error bars show the same for the fluctuations measure in $\Reh = u_0 \ell_0/\nu$ over time. Lines show fits of $\gamma_\texp \propto \Reh^{1/2}$ to subsonic points (solid line) and $\gamma_\texp \propto \Reh^{1/3}$ to supersonic points (dashed line), which match predictions from \cite{schober2012small, schober2012magnetic}, indicating that the viscous scale is the engine for growth in the kinematic regime.}
        \label{fig:gamma_exp_scaling}
    \end{figure}
    
    In Fig.~\ref{fig:gamma_exp_scaling} we show our inferred kinematic-phase growth rates, $\gamma_\texp t_0$, for each plasma configuration. Consistent with both theoretical expectation and prior work, we find that the data follow $\gamma_\texp \propto \Reh^{1/2}$ for $\Mach \lesssim 1$ and $\gamma_\texp \propto \Reh^{1/3}$ for $\Mach > 1$ simulations. Both scalings align with the expectation that magnetic growth is regulated by the stretching on the viscous scale of the hydrodynamical cascade, $\gamma_\texp \sim u_\nu/\ell_\nu \sim \Reh^{(1-\vartheta)/(1+\vartheta)}$, with $\vartheta = 1/2$ for Kolmogorov-like, and $\vartheta = 1/3$ for Burgers-like turbulence \citep{kulsrud1992spectrum, schekochihin2002model, schober2012magnetic}. We find that the growth rate transitions sharply between the two regimes, \ie as soon as the turbulence becomes even mildly supersonic the growth rate follows $\gamma_\texp \propto \Reh^{1/3}$.
    
    While the scaling of $\gamma_\texp$ with $\Reh$ in our simulations is in agreement with theoretical expectations, the proportionality constants we measure, of order $\sim 1/100$, are systematically smaller than expected. A simple least-squares fit of the dimensionless kinematic growth rate, $\gamma_\texp t_0$, as a function of $\Reh$, gives $(3.22 \pm 0.06) \times 10^{-2} \,\Reh^{1/2}$ for our $\Mach \lesssim 1$ simulations, and $(4.48 \pm 0.06) \times 10^{-2} \,\Reh^{1/3}$ for our $\Mach > 1$ simulations. These fits indicate that the kinematic dynamo only converts $\sim 1/100$ of the hydrodynamical energy flux, $\varepsilon$, into magnetic energy. We show in the next section that this same conversion efficiency also holds in the nonlinear dynamo phase. These values are significantly smaller than predicted for $\Prm \gg 1$ flows ($37/36$ for Kolmogorov flows, and $11/60$ for Burgers \citep{schober2012small, schober2012magnetic}) and instead align more closely with predictions for $\Prm \ll 1$ flows, where $0.03$ is expected for Kolmogorov, and $5\times10^{-3}$ for Burgers \citep{schober2012small, schober2012magnetic, bovino2013turbulent}.

\subsection{Nonlinear dynamo growth}

    \begin{figure}
        \centering
        \includegraphics[width=\linewidth]{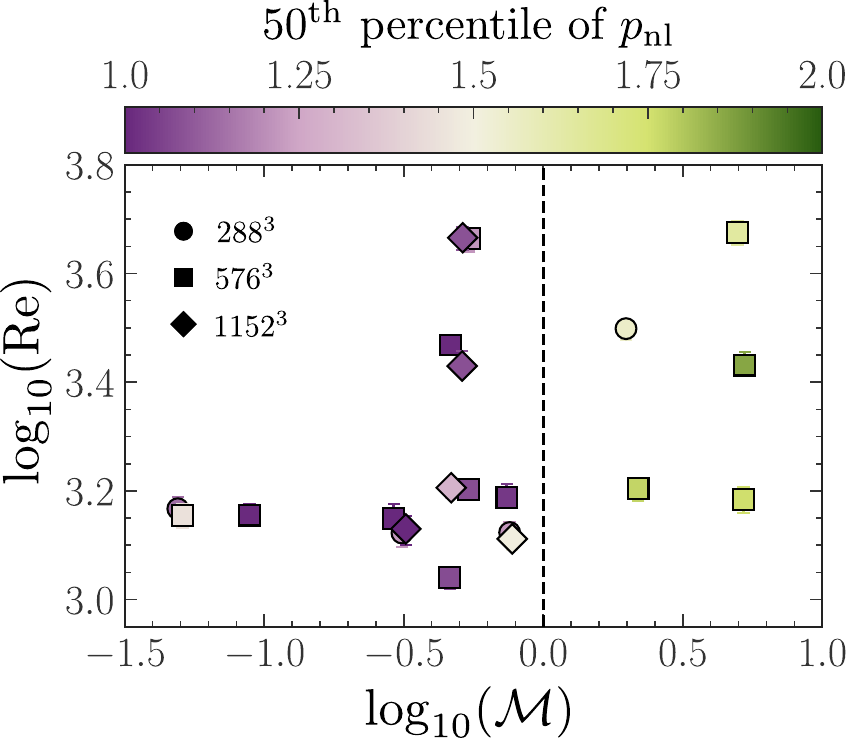}
        \caption{Secular dynamo growth exponent $p_\tnl$, measured during the nonlinear phase, as a function of $\Mach$ and $\Reh$ for each plasma configuration. For $\Mach \lesssim 1$ simulations we unanimously recover linear growth ($p_\tnl \approx 1$; purple), and quadratic growth ($p_\tnl \approx 2$; green) for all $\Mach > 1$ simulations.}
        \label{fig:nl_exponent}
    \end{figure}

    \begin{figure}
        \centering
        \includegraphics[width=\linewidth]{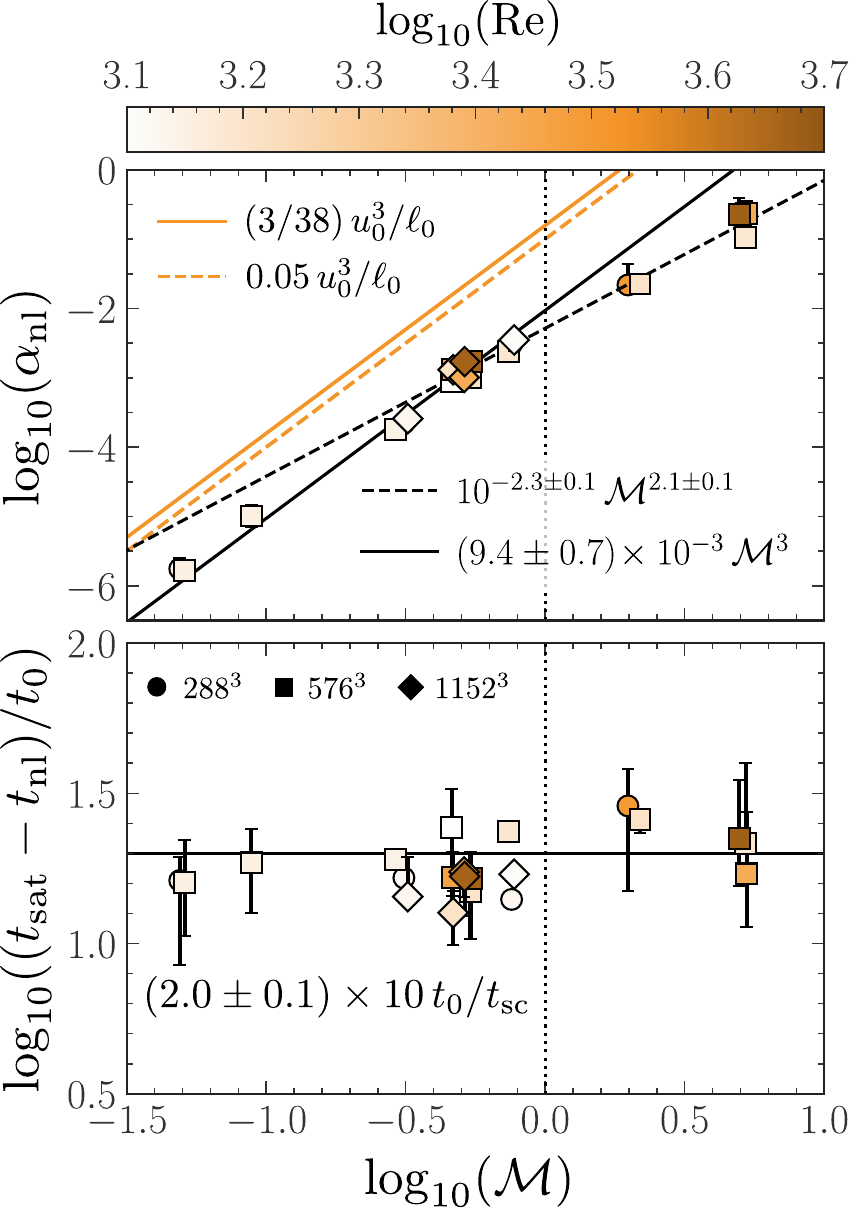}
        \caption{Nonlinear growth coefficient $\alpha_\tnl$ (top panel) and duration of the nonlinear phase normalized by $t_0$ (bottom panel), as a function of $\Mach$, and colored by $\Reh$. Points show median posterior values as derived from the MCMC, while error bars show the 16th to 84th percentile range, although in some cases the error bars are not visible because they are smaller than the plot markers. For $\Mach \lesssim 1$ we find $\alpha_\tnl \propto \Mach^3$ (black solid line), consistent with turbulent energy-flux-regulated models of nonlinear growth: $0.05 u_0^3/\ell_0$ from \citep{beresnyak2012universal} (dot-dashed orange) and $(3/38)u_0^3/\ell_0$ from \citep{xu2016turbulent} (dotted orange). By contrast, for $\Mach > 1$, we find $\alpha_\tnl \propto \Mach^2$ (black dashed), which is shallower than the flux-regulated prediction. Critically, $\alpha_\tnl$ in both $\Mach$-regimes becomes independent of $\Reh$, and the cascade itself becomes the dynamo engine, where we find that only a roughly constant fraction of the turbulent kinetic energy flux is converted into magnetic energy, \viz $(dE_\tmag/dt)/\varepsilon \approx 1/100$. Across all simulations the nonlinear phase persists for a universal $t_\tsat - t_\tnl \approx 20 t_0$ (solid black line in lower panel) duration, invariant to all plasma parameters explored in this study.}
        \label{fig:nl_scalings}
    \end{figure}

    Fig.~\ref{fig:nl_exponent} shows our measured growth-exponent, $p_\tnl$, for $E_\tmag$ during the nonlinear phase, and Fig.~\ref{fig:nl_scalings} shows both the growth-efficiency coefficient, $\alpha_\tnl$, in the top panel, and duration, $t_\tsat - t_\tnl$, of the nonlinear phase in the bottom panel. For our $\Mach \lesssim 1$ simulations, we find that growth is close to linear-in-time, $p_\tnl \approx 1$, and a least-squares fit to the median results shows that the growth coefficient is $\alpha_\tnl = (9.4 \pm 0.7) \times 10^{-3} \Mach^3 \approx 5\times 10^{-3}\varepsilon$, where $\varepsilon \sim u_0^3 /\ell_0 = 2 \Mach^3$ is the hydrodynamic energy flux in our simulations. Critically, the growth rate is independent of $\Reh$, suggesting that, even though we have explored only a finite range $\Reh \in [10^3, 5\times 10^3]$, the results are likely to be valid for all turbulent flows (where $\Reh > \Reh_\tcrit \approx 100$; see \citet{kriel2022fundamental}). This is qualitatively consistent with previous works that have shown only a small, fixed fraction of $\varepsilon$ is converted into magnetic energy \citep{cho2009growth, beresnyak2012universal}, but it has never been measured with this accuracy and precision, across a broad range of plasma parameters. By comparison, \citet{beresnyak2012universal} measured $\alpha_\tnl /\varepsilon \approx 0.05$ (compared to our $9.4 \pm 0.7 \times 10^{-3} \sim 10^{-2}$) from ensemble averaged $\Mach \ll 1$, $\Prm = 1$ simulations, and \citet{xu2016turbulent} predicted $\alpha_\tnl /\varepsilon = 3/38$ for $\Mach \ll 1$, $\Prm \gg 1$ SSDs. We show both of these predictions in Fig.~\ref{fig:nl_scalings}, and while our measured efficiency is a factor of $\approx 10$ lower than both predictions, as was also the case for the growth rate measured during the kinematic phase, the difference, at least compared with \citet{xu2016turbulent}, can be explained by our finite $\Prm = 1$ simulations. Regardless, the overall results are consistent with the phenomenology of inefficient conversion of turbulent to magnetic energy in the nonlinear phase.
    
    All of our $\Mach > 1$ simulations yield $p_\tnl \approx 2$, consistent with quadratic-in-time growth \footnote{A complementary model-comparison analysis (see End Matter) confirms that subsonic results are best described by linear growth, while quadratic growth captures the supersonic regime.} of $E_\tmag$, aligned with the model of \citet{schleicher2013small}, which predicts that $E_\tmag(t) \sim E_\tkin^{1/2} (t/t_0)^2$ for Burgers-like turbulence. Indeed, as predicted by \citet{schleicher2013small}, this breaks the universality of $p_\tnl$ across different $\Mach$ regimes. Moreover, unlike the subsonic case, $\varepsilon$ no longer scales with $\Mach^3$, but instead follows a shallower trend $\alpha_\tnl \propto \varepsilon \propto \Mach^2$. Empirically, the trends we measure can be understood if the conversion of $E_{\tkin}$ into $E_{\tmag}$ is modified by the presence of acoustic modes and shocks, \ie if the flux is transferred on $t_\tsc$ timescales, due to a fraction of the energy being directly deposited into shock heating, $\varepsilon \sim u^2/t_{\tsc} \sim u^2 c_s / \ell \sim \Mach^2$, which acts to reduce the available hydrodynamical flux by a factor of $\Mach^{-1}$. This is consistent with the idea that in the compressible regime, some fraction of the $E_{\tkin}$ fills compressible mode degrees of freedom, which do not contribute to irreversible field amplification \citep{Federrath2011_supersonic_dynamo, Federrath2014_supersonic_dynamo, seta2021saturation, sur2024role, beattie2025taking}.
    
    Despite the differences in scaling and $\Mach$-dependence between the $\Mach \lesssim 1$ and $\Mach > 1$ regimes, we find two aspects of the nonlinear SSD that are universal across plasma parameters and $\Mach$-regimes: firstly, the proportionality coefficient, which is associated with the nonlinear SSD efficiency, is universal across all $\Mach$ and $\Reh$, with $\varepsilon / 100$ efficiency consistent with what was found previously \citep{beresnyak2012universal}, even though $p_\tnl$ is not always unity. And secondly, across the full parameter space, we find that the nonlinear phase persists for $t_\tsat - t_\tnl \approx (20 \pm 1) \,t_0$, independent of $\Mach$ or $\Reh$ (see Fig.~\ref{fig:nl_scalings}, bottom panel), making the duration in the nonlinear dynamo also universal. This suggests that once the nonlinear phase begins, $E_\tmag \approx E_{\tkin}(\ell_\nu)$, the system saturates after $\approx 20 t_0$, regardless of flow compressibility. This universal nonlinear window is visually apparent in both the bottom and inset panels of Fig.~\ref{fig:time_series}, where the nonlinear phase occupies a comparable interval when time is expressed as $t/t_0$ or $\log_{10}(t/t_\tsc)$. This is likely because, although $\varepsilon$ and $\alpha_\tnl$ have different $\Mach$-dependencies, their ratio remains $\alpha_\tnl / \varepsilon \approx 1/100$ for all dynamo (both kinematic and nonlinear) and flow regimes (sub- and supersonic) (Fig.~\ref{fig:nl_scalings}, top panel). The implication is that while the algebraic order and $\Mach$ dependence of $E_\tmag$ growth depends on the turbulent regime, the path to saturation is set by a robust and universal dynamical clock.

\section{Discussion and Conclusions}

    In this study we present a detailed statistical analysis of small-scale dynamo (SSD) growth across a range of hydrodynamic Reynolds, $\Reh$, and sonic Mach numbers $\Mach$, focusing on the poorly-explored nonlinear phase of field growth. In order to access this regime, we (1) simulate many statistical realizations of each set of plasma parameters, allowing our measurements to become less sensitive to large statistical fluctuations that make this phase challenging to explore, and (2) use a hierarchical Bayesian model fitting technique that allows us to precisely model the dynamo phase transitions. We find that in this regime the growth rate depends on $\Mach$, but is liberated from any visco-resistive dynamics. $\Mach < 1$ nonlinear SSDs grow $\propto t$ with an efficiency of order $\approx 1/100$ of the hydrodynamic energy flux from the turbulence cascade, consistent with the phenomenological models with the incompressible nonlinear SSD \citep{schekochihin2002model, xu2016turbulent}. Further, $\Mach > 1$ dynamos grow $\propto t^2$, predicted by existing theories \citep{schober2012magnetic, schleicher2013small}, and with the same SSD efficiency $\approx 1/100$, making both the sub- and supersonic nonlinear dynamos universally inefficient in their parasitism of the hydrodynamical cascade \citep{beresnyak2012universal}.
    
    Following from the universal efficiency, we find that the nonlinear phase has an approximately universal duration of $\approx 20t_0$, where $t_0$ is the outer-scale turbulence turnover time, independent of both the $\Reh$ and $\Mach$. Consequently, the time it takes systems to evolve through the nonlinear phase and reach $E_{\tmag}$ saturation is always roughly an order of magnitude longer than the time it takes systems to reach mechanical equilibrium $\sim t_0$, but no more. This implies that most observed astrophysical systems, which persist for many $t_0$, are likely to have reached SSD saturation; only those that have undergone large perturbations very recently, or that are very dynamically young, are likely to be in the nonlinear phase. Why the efficiency of the nonlinear SSD is $\sim 1/100$, and the characteristic duration of the phase $\approx 20 t_0$, rather than any other values, remains an open theoretical question.
    
    We note that our study has focused on $\Prm = 1$ plasmas, while many hot astrophysical plasmas are $\Prm \gg 1$ ($\Prm \propto T^4/n_e$, where $T$ is the plasma temperature and $n_e$ is the electron number density). Our choice of $\Prm = 1$ is motivated by the fact that in this study we target the nonlinear phase that begins once $E_\tmag \approx E_\tkin$ on the viscous scale, and the backreaction that operates on dynamically important (viscous and larger) scales. In $\Prm \gg 1$ systems, theory predicts that the nonlinear phase studied here may be preceded by another nonlinear, secular dynamo phase, where the magnetic field traverses the sub-viscous range of scales before reaching the viscous scale \citep{schekochihin2002model}. Our results therefore provide a useful anchor for future studies that explore this earlier stage of nonlinear evolution.
    
    Finally, we note that our $\Mach > 1$ simulations do not resolve a $u_{\ell_s} = c_s$ sonic scale, where the velocity spectrum transitions between a subsonic and supersonic cascade \citep{Federrath2021_the_sonic_scale, Beattie2025_nature_astronomy}. It may be that the nonlinear SSD transitions smoothly between the $E_\tmag \sim \Mach^3 t$ and $E_\tmag \sim \Mach^2 t^2$ growth phase, as in Fig.~\ref{fig:nl_scalings}, which may be relevant for $\Mach \gg 1$, $\Reh \gg 1$ astrophysical plasmas. We leave the study of this regime for future, extremely high-resolution nonlinear SSD investigations.

\begin{acknowledgments}
    We are deeply grateful to Christoph Federrath for allowing us to use his version of the \textsc{flash} code \citep{Federrath2021_the_sonic_scale, Federrath2022_TurbGen}, which enabled this project. We also thank Tomasz Rozanski and Cameron Van Eck for many helpful discussions surrounding our Bayesian analysis. N.~K. and M.~R.~.K acknowledge support from the Australian Research Council through Laureate Fellowship award FL220100020. This research was undertaken with the assistance of resources from the National Computational Infrastructure (NCI Australia), an NCRIS enabled capability supported by the Australian Government, through award jh2. J.~R.~B. acknowledges funding from the Natural Sciences and Engineering Research Council of Canada (NSERC, funding reference number 568580); support from NSF Award 2206756; and high-performance computing resources provided by the Leibniz Rechenzentrum and the Gauss Center for Supercomputing (grants pn76gi, pr73fi, and pn76ga).

\end{acknowledgments}

\section{End Matter}

\subsection{Simulation Setup}

    For all simulations we solve the compressible set of non-ideal (visco-resistive) magnetohydrodynamical fluid equations,
    \begin{align}
        \priv{\rho}{t}
            + \vect{\nabla}\cdot(\rho \vect{u})
            &= 0
            , \\
        \priv{\rho\vect{u}}{t}
            + \vect{\nabla}\cdot\bigg[
                \rho\vect{u}\otimes\vect{u}
                - \frac{1}{4\pi}\vect{b}\otimes\vect{b}
                \nquad[5]\nonumber \\
                + \rbrac{c_\ts^2\rho + \frac{b^2}{8\pi}} \tensor{\bm{I}}
                - 2\nu\rho \tensor{\bm{S}}
            \bigg] &= \rho\vect{f}
            , \\
        \priv{\vect{b}}{t}
            + \vect{\nabla}\cdot\rbrac{\vect{u}\otimes\vect{b}-\vect{b}\otimes\vect{u}} 
            &= \eta \vect{\nabla}^2\vect{b}
            , \\
        \vect{\nabla}\cdot\vect{b}
            &= 0
            ,
    \end{align}
    where $\rho$ is gas density, $\vect{u}$ is the gas velocity, $\vect{b}$ is the the magnetic field, $c_\ts$ is sound speed, $\nu$ is the kinematic viscosity, $\eta$ is the magnetic resistivity, $\tensor{\bm{I}} = \delta^i_j$ is the identity tensor. We model our viscosity via the traceless strain rate tensor,
    \begin{align}
        \tensor{\bm{S}}
            = \frac{1}{2} \rbrac{\vect{\nabla}\otimes\vect{u} + \rbrac{\vect{\nabla}\otimes\vect{u}}^T}
            - \frac{1}{3} \rbrac{\vect{\nabla}\cdot\vect{u}} \tensor{\bm{I}}
            ,
    \end{align}
    where $\otimes$ is the tensor product $\vect{\nabla}\otimes\vect{u} \equiv \pp^i u_j$. Our simulations span a broad range of $\Mach$ and $\Reh$, and are run at a range of resolutions; see Section~\ref{sec:methods:sims} for details and Table~\ref{table:summary} for a full list. The forcing function $\vect{f}$ is Gaussian-random and the phases are evolved in time using the \textsc{turbgen} implementation \citep{Federrath2022_TurbGen} of an Ornstein-Uhlenbeck process, where the correlation time sets to outer-scale of the turbulence, $t_0$. We set the injection to peak on $\ell_0 = L/2$, and tune the forcing amplitude so that the velocity field on $\ell_0$ stays within 5\% of a chosen target value during the kinematic phase; we explore $\Mach \in [5\times 10^{-2}, 5]$. 

    \begin{table*}
\renewcommand{\arraystretch}{1.25}
\setlength{\tabcolsep}{3.5pt}
\caption{Summary of simulation configurations. Columns 1–3 list key plasma parameters, column 4 the number of instances at a particular resolution, and columns 5–8 the dimensionless parameters inferred from our MCMC fitting routine for the ensemble-averaged runs at the respective resolution. Note, we report columns 5-8 in dimensionless units.}
\label{table:summary}
\begin{center}
  \begin{tabular}{
        c c c c
        c c c c
    }
    \hline\hline
    $\Mach$
    & $\Reh$
    & $\nu\,t_0/\ell_0^2$
    & Runs
    & $\gamma_\texp t_0$
    & $\alpha_\tnl$
    & $p_\tnl$
    & $(t_\tsat - t_\tnl) / t_0$
    \\[0.4em]
    (1) & (2) & (3) & (4) & (5) & (6) & (7) & (8) \\
\hline
\hline

0.05 & 1500 & $1.7\times 10^{-5}$ & 5\,$\times$\,288 & $1.1_{-0.1}^{+0.1}$ & $\left(1.8_{-0.4}^{+0.7}\right)\times 10^{-6}$ & $1.1_{-0.1}^{+0.8}$ & $16.2_{-7.7}^{+3.1}$ \\

0.05 & 1500 & $1.7\times 10^{-5}$ & 5\,$\times$\,576 & $1.2_{-0.1}^{+0.1}$ & $\left(1.7_{-0.5}^{+0.6}\right)\times 10^{-6}$ & $1.4_{-0.4}^{+0.4}$ & $16.0_{-5.4}^{+6.1}$ \\

0.1 & 1500 & $3.3\times 10^{-5}$ & 5\,$\times$\,576 & $1.26_{-0.07}^{+0.05}$ & $\left(1.0_{-0.3}^{+0.5}\right)\times 10^{-5}$ & $1.0_{-0.1}^{+0.7}$ & $18.5_{-5.9}^{+5.5}$ \\

0.3 & 1500 & $1\times 10^{-4}$ & 1\,$\times$\,288 & $1.3_{-0.1}^{+0.1}$ & $\left(2.22_{-0.02}^{+0.02}\right)\times 10^{-4}$ & $1.22_{-0.03}^{+0.03}$ & $16.5_{-0.2}^{+0.3}$ \\

0.3 & 1500 & $1\times 10^{-4}$ & 1\,$\times$\,576 & $1.2_{-0.1}^{+0.1}$ & $\left(1.82_{-0.01}^{+0.01}\right)\times 10^{-4}$ & $1.0_{-0.1}^{+0.1}$ & $19.0_{-0.1}^{+0.1}$ \\

0.3 & 1500 & $1\times 10^{-4}$ & 3\,$\times$\,1152 & $1.30_{-0.02}^{+0.01}$ & $\left(2.6_{-0.4}^{+0.4}\right)\times 10^{-4}$ & $1.00_{-0.10}^{+0.01}$ & $14.3_{-0.3}^{+5.1}$ \\

0.5 & 1000 & $2.5\times 10^{-4}$ & 9\,$\times$\,576 & $\left(9.5_{-0.4}^{+0.3}\right)\times 10^{-1}$ & $\left(9_{-2}^{+5}\right)\times 10^{-4}$ & $1.1_{-0.1}^{+0.9}$ & $24.3_{-11.8}^{+8.5}$ \\

0.5 & 1500 & $1.7\times 10^{-4}$ & 6\,$\times$\,576 & $1.3_{-0.1}^{+0.2}$ & $\left(0.7_{-0.6}^{+0.2}\right)\times 10^{-3}$ & $1.1_{-0.1}^{+0.5}$ & $14.9_{-4.5}^{+5.4}$ \\

0.5 & 1500 & $1.7\times 10^{-4}$ & 3\,$\times$\,1152 & $1.26_{-0.04}^{+0.06}$ & $\left(1.3_{-0.1}^{+0.4}\right)\times 10^{-3}$ & $1.3_{-0.3}^{+0.6}$ & $12.7_{-2.8}^{+2.3}$ \\

0.5 & 3000 & $8.3\times 10^{-5}$ & 9\,$\times$\,576 & $1.8_{-0.2}^{+0.1}$ & $\left(1.3_{-0.1}^{+0.2}\right)\times 10^{-3}$ & $1.0_{-0.1}^{+0.2}$ & $16.6_{-2.2}^{+3.6}$ \\

0.5 & 3000 & $8.3\times 10^{-5}$ & 3\,$\times$\,1152 & $1.9_{-0.1}^{+0.2}$ & $\left(1.0_{-0.1}^{+0.2}\right)\times 10^{-3}$ & $1.08_{-0.08}^{+0.10}$ & $17.3_{-3.0}^{+0.9}$ \\

0.5 & 5000 & $5\times 10^{-5}$ & 1\,$\times$\,576 & $2.42_{-0.01}^{+0.01}$ & $\left(1.73_{-0.01}^{+0.02}\right)\times 10^{-3}$ & $1.22_{-0.03}^{+0.03}$ & $16.5_{-0.2}^{+0.2}$ \\

0.5 & 5000 & $5\times 10^{-5}$ & 5\,$\times$\,1152 & $2.4_{-0.3}^{+0.2}$ & $\left(1.7_{-0.3}^{+0.4}\right)\times 10^{-3}$ & $1.1_{-0.1}^{+0.2}$ & $16.7_{-4.3}^{+1.3}$ \\

0.8 & 1500 & $2.7\times 10^{-4}$ & 1\,$\times$\,288 & $1.1_{-0.1}^{+0.1}$ & $\left(3.06_{-0.04}^{+0.05}\right)\times 10^{-3}$ & $1.22_{-0.02}^{+0.02}$ & $14.0_{-0.3}^{+0.2}$ \\

0.8 & 1500 & $2.7\times 10^{-4}$ & 1\,$\times$\,576 & $\left(9.9_{-0.1}^{+0.1}\right)\times 10^{-1}$ & $\left(2.40_{-0.01}^{+0.01}\right)\times 10^{-3}$ & $1.03_{-0.01}^{+0.01}$ & $23.6_{-0.1}^{+0.1}$ \\

0.8 & 1500 & $2.7\times 10^{-4}$ & 3\,$\times$\,1152 & $1.1_{-0.1}^{+0.1}$ & $\left(4_{-1}^{+1}\right)\times 10^{-3}$ & $1.5_{-0.5}^{+0.4}$ & $17.0_{-1.0}^{+0.3}$ \\

2.0 & 1500 & $6.7\times 10^{-4}$ & 5\,$\times$\,576 & $\left(5.1_{-0.6}^{+0.8}\right)\times 10^{-1}$ & $\left(2.3_{-0.4}^{+0.2}\right)\times 10^{-2}$ & $1.8_{-0.8}^{+0.2}$ & $25.9_{-2.6}^{+2.2}$ \\

2.0 & 3000 & $3.3\times 10^{-4}$ & 5\,$\times$\,288 & $\left(7_{-1}^{+1}\right)\times 10^{-1}$ & $\left(2_{-1}^{+2}\right)\times 10^{-2}$ & $1.6_{-0.5}^{+0.3}$ & $28.7_{-13.8}^{+9.3}$ \\

5.0 & 1500 & $1.7\times 10^{-3}$ & 5\,$\times$\,576 & $\left(5.3_{-0.5}^{+0.3}\right)\times 10^{-1}$ & $\left(1.1_{-0.3}^{+0.5}\right)\times 10^{-1}$ & $1.8_{-0.6}^{+0.2}$ & $21.5_{-1.5}^{+18.4}$ \\

5.0 & 3000 & $8.3\times 10^{-4}$ & 5\,$\times$\,576 & $\left(6.7_{-0.3}^{+0.2}\right)\times 10^{-1}$ & $\left(2_{-1}^{+1}\right)\times 10^{-1}$ & $1.9_{-0.8}^{+0.1}$ & $17.0_{-5.7}^{+10.4}$ \\

5.0 & 5000 & $5\times 10^{-4}$ & 5\,$\times$\,576 & $\left(7.3_{-0.3}^{+0.7}\right)\times 10^{-1}$ & $\left(2_{-1}^{+2}\right)\times 10^{-1}$ & $1.6_{-0.4}^{+0.4}$ & $22.3_{-6.8}^{+12.8}$ \\
    
\hline
\hline
    \end{tabular}
\end{center}
\end{table*}

\subsection{MCMC parameters}

    We carried out all fits using the \texttt{emcee} ensemble sampler \citep{Foreman-Mackey13a}. For each parameter vector we employed $10$ walkers per free parameter, each evolved for $10^4$ steps, where the first $3\times 10^3$ steps were discarded as burn-in, with no thinning applied. In the first stage, walkers were initialized with small, $1\%$ Gaussian scatter around the prior ranges, while in the second stage we initialized them with a $1\%$ Gaussian scatter relative to the median parameters inferred from stage~1. Convergence was verified by monitoring the integrated auto-correlation time of the chains, by inspecting the stability of the posterior distributions, and by checking that the median acceptance fraction across walkers lay within the recommended range of $0.2$-$0.5$. As discussed in the main text, all posteriors are based on the combined post-burn-in samples, and the measurements reported are of percentiles over the combined ensembles.

\subsection{Model comparison with AIC}

    To further test the robustness of our inference, we employed the Akaike Information Criterion (AIC). For each dataset $i$ and candidate model \mbox{$j \in \mathcal{N} \equiv \{p_\tnl=1, \,p_\tnl=2\}$}, we calculate
    \begin{align}
        \mathrm{AIC}_{ij}
            = 2k - 2\ln \mathcal{L}(d_i \mid \bm{\hat{\theta}}_j) ,
    \end{align}
    where $d_i$ denotes a unique plasma configuration instance, $\bm{\hat{\theta}}_j$ are the maximum-likelihood parameters for model $j$, and $k = 5$ is the number of free parameters (which is the same for both models). Model comparison is then based on the relative AIC weights,
    \begin{align}
        w_{ij}
            = \frac{
                \exp\!\rbrac{-\tfrac{1}{2} \Delta_{ij}}
            }{
                \sum_{n \in \mathcal{N}} \exp\!\rbrac{-\tfrac{1}{2} \Delta_{in}}
            } ,
    \end{align}
    with
    \begin{align}
        \Delta_{ij}
            = \mathrm{AIC}_{ij} - \min_{n \in \mathcal{N}} \mathrm{AIC}_{in} .
    \end{align}
    The weights $w_{ij}$ quantify the probability that model $j$ is preferred for dataset $i$. In practice, for many datasets the weight of the favored model is numerically $w \approx 1$ while the alternative has $w \approx 0$ (to machine precision). This outcome is expected when the number of independent data points is large, and gives us confidence that the nonlinear phase has been sufficiently resolved to independently constrain the dynamics in this transitionary regime.  
    
    Among the $\Mach \lesssim 1$ simulations we find that most cases ($19/30 \approx 63\%$) favor the linear model ($p_\tnl = 1$), while the quadratic model ($p_\tnl = 2$) is preferred ($29/40 \approx 73\%$) by the $\Mach > 1$ simulations. These findings independently confirm the dichotomy in nonlinear growth behavior shown in Figs.~\ref{fig:nl_exponent} and \ref{fig:nl_scalings}, and the relative fractions are consistent with those inferred from our MCMC fits within their statistical uncertainties.

\bibliography{refs}

\end{document}